\documentclass[showpacs,aps,pre,superscriptaddress,twocolumn]{revtex4}
\usepackage{graphicx}
\usepackage{amsfonts}
\usepackage{color}

\begin{document}
\title{Compactons in Nonlinear Schr\"odinger Lattices with Strong Nonlinearity Management}
\author{F.Kh. Abdullaev}
\affiliation{CFTC, Universidade de Lisboa,   Av. Prof.Gama Pinto 2, Lisboa 1649-003, Portugal}
\author{P.G. Kevrekidis}
\affiliation{Department of Mathematics and Statistics, University
of Massachusetts, Amherst, MA 01003}
\author{M. Salerno}
\affiliation{Dipartimento di Fisica ``E.R. Caianiello'', CNISM  and INFN -
Gruppo Collegato di Salerno,
Universit\`a di Salerno, Via Ponte don Melillo, 84084 Fisciano (SA), Italy}

\begin{abstract}
The existence of compactons in the discrete nonlinear Schr\"odinger
equation in the presence of
fast periodic time modulations of the nonlinearity is demonstrated.
In the averaged DNLS equation the resulting effective inter-well tunneling depends
on
modulation parameters {\it and} on the field amplitude.
This  introduces  nonlinear dispersion in the system and can lead to a
prototypical realization of
single- or multi-site stable
discrete  compactons in nonlinear optical waveguide and BEC arrays.
These structures can dynamically arise out of Gaussian or compactly
supported initial data.
\end{abstract}
\maketitle
\newpage

\noindent {\it Introduction}.
One of the most remarkable phenomena occurring in nonlinear lattices is the existence  of {\it discrete breathers} which arise from the interplay between discreteness, dispersion and nonlinearity~\cite{Flach}.
These  excitations are quite  generic in nonlinear lattices with usual (e.g. linear) dispersion and have typical spatial profiles with exponential tails.
In presence of {\it nonlinear dispersion} these excitations  (as well as their continuous counterparts) may acquire spatial profiles with compact support  and for this reason they are known as {\it compactons}~\cite{rosenau}. Unlike other nonlinear excitations, compactons (having  no tails)
cannot interact with each other until they are in contact, this being an attractive feature for
potential applications. Similarly to discrete breathers, compactons are
intrinsically localized and robust excitations. The lack of  exponential tails
is a consequence of the nonlinear dispersive interactions which permit the
vanishing of the intersite tunneling at compacton edges.  The difficulty of
implementing this condition in physical  contexts has restricted until now
investigations mainly to the mathematical side. The development of management
techniques for soliton control, however,  can rapidly change the situation.

Periodic management of parameters of nonlinear systems has been shown to be  an effective technique for the generation of solitons with new types of properties~\cite{Malomed_book}.
Examples of the management technique in continuous systems are the
dispersion management of solitons in optical fibers which allows to improve communication capacities~\cite{disp_man}, and the nonlinearity management of 2D and 3D Bose-Einstein
condensates (BEC) or optically layered media which provides partial stabilization against
collapse in the case of attractive interatomic interactions~\cite{Saito}.
In discrete systems the diffraction management technique was used to generate spatial discrete solitons with novel properties~\cite{AM,Garanovich} which have
been  observed in experiments~\cite{Garanovich}.
The resonant spreading and steering of discrete solitons  in arrays of waveguides, induced by nonlinearity management was also investigated~\cite{Assanto}.
To date, the nonlinear management technique {\it for nonlinear lattices} has been considered only in the limit of weak modulations of the nonlinearity~\cite{ATMK, Kartashov05}.
The inter-well tunneling suppression has been discussed in~\cite{HolthausBH} for the Bose-Hubbard chain with time periodic  ramp potential and in~\cite{Gong} for a two-sites  Bose-Hubbard model with modulated in time interactions. In both  cases the tunneling suppression was uniform in the system and no apparent link with compacton formation  was established.  The phenomenon has also been  recently observed in experiments of light propagation in waveguide arrays~\cite{Szameit} and in BEC's in strongly driven optical lattices~\cite{arimondo}.

The aim of the present Letter is to demonstrate the existence of stable compacton excitations in the discrete nonlinear Schr\"odinger (DNLS) system subjected to  {\it strong nonlinearity management} (SNLM), e.g. to fast periodic time variations of the nonlinearity.
To that effect, we use an averaged DNLS Hamiltonian system to show that in the SNLM limit the inter-well tunneling can be totally suppressed for field amplitudes matching  zeros of the Bessel function,
introducing effective nonlinear dispersion  which leads to  compacton
formation. We show that these compact  structures not only
exist in single and multi-site realizations but they generically are
 structurally and
dynamically stable and can be generated from general classes
of initial conditions. These results  should enable the observation of discrete
compactons in  BEC and in nonlinear optical systems, both being
described by the discrete NLS equation.

\noindent {\it Theory.} Consider  the following lattice Hamiltonian
\begin{equation}
H= - \sum_{n} \{\kappa  (u_n  u_{n+1}^* +
u_{n+1} u_{n}^*) + \frac 12 (\gamma_0+\gamma(t)) |u_n|^4 \},
\label{dnlsH}
\end{equation}
with the coupling constant $\kappa$  quantifying the tunneling between adjacent sites (wells), $\gamma_0$  denoting  the
onsite constant nonlinearity  and $\gamma(t)$ representing the time-dependent modulation. In the following we assume
a strong management case with
$\gamma(t)$ being a periodic, e.g.  $\gamma(t)=\gamma(t+T)$,  and rapidly varying function. As a prototypical example, we use
$\gamma (t)=\frac {\gamma_1}\varepsilon \cos(\Omega \tau)$,
with $\gamma_1 \sim O(1)$, $\varepsilon \ll 1$, $\tau=t/\varepsilon$ denoting the fast time variable and $T=2 \pi/\Omega$ the period.
The dynamical system  associated with (\ref{dnlsH}) is  the well known DNLS equation \cite{pgk_book}
\begin{equation}\label{dnls}
i \dot{u}_{n} + \kappa(u_{n+1} + u_{n-1}) + (\gamma_0 + \gamma(t))|u_n|^2 u_n = 0,
\end{equation}
which serves, under suitable conditions \cite{ABKS},
as a model for the  dynamics of  BEC in optical lattices
subjected to SNLM (through varying the interatomic
scattering length  by
external time-dependent  magnetic fields via a Feshbach resonance),
as well as for
light propagation in optical waveguide arrays (here the
evolution variable is the propagation distance and the SNLM
consists of periodic space variations of the Kerr nonlinearity).

The existence of compacton solutions can be inferred from the fact that the averaged DNLS Hamiltonian (averaged with respect to the fast time $\tau$),  coincides with the original time independent Hamiltonian except for a rescaling of the coupling constant which depends on the Bessel function of the field amplitude.
To show this, it is convenient to  perform the transformation \cite{JKP}
$u_n(t) = v_n(t)e^{i\Gamma |v_n(t)|^2}$ with $\Gamma =
\frac{1}{\epsilon}\int_0^t dt\ \gamma (\tau) = \gamma_1 \Omega^{-1} \sin(\Omega \tau)$,
which allows to rewrite Eq. (\ref{dnls}) as
\begin{equation}
i \dot v_n = \Gamma v_n (|v_n|^2)_t - \kappa X -\gamma_0 |v_n|^2 v_n ,
\label{ms1}
\end{equation}
with $X= v_{n+1} e^{i \Gamma \theta_{+}} + v_{n-1} e^{i \Gamma \theta_{-}}$ and $\theta_{\pm}= |v_{n\pm 1}|^2- |v_n|^2$. On the other hand,  $(i |v_n|^2)_t= i (\dot v_n v_n^*+ v_n \dot v_n ^*)=i \kappa (v_n^* X - v_n X^*)$, with the star denoting the complex conjugation. Substituting this expression  into Eq. (\ref{ms1})  and averaging
the resulting equation over the period $T$ of the rapid modulation, we obtain
\begin{equation}
i \dot v_n = i \kappa |v_n|^2 \langle \Gamma X \rangle - i \kappa v_n^2 \langle \Gamma X^* \rangle - \kappa \langle X \rangle -\gamma_0 |v_n|^2 v_n ,
\label{ms2}
\end{equation}
with $\langle \cdot \rangle \equiv \frac 1T \int_0^T ( \cdot ) d\tau$ denoting the fast time average. The averaged terms in Eq. (\ref{ms2}) can be calculated by means of the elementary integrals $\langle e^{\pm i \Gamma \theta_\pm} \rangle= \alpha J_0 (\alpha \theta_\pm)$,
$\; \langle \Gamma e^{\pm i \Gamma \theta_\pm} \rangle=\pm i \alpha J_1 (\alpha \theta_\pm)$,  with $J_i$ being Bessel functions of order $i=0,1$ and  $\alpha = \gamma_1/\Omega$, thus giving
\begin{eqnarray}
i \dot{v}_{n}  =  - \alpha \kappa v_n [(v_{n+1}v_n^{\ast} + v_{n+1}^{\ast} v_n) J_1(\alpha\theta_{+})+\nonumber\\
(v_{n-1}v_n^{\ast} + v_{n-1}^{\ast} v_n)J_1(\alpha\theta_{-})] - \nonumber \\
\kappa[v_{n+1} J_0(\alpha\theta_{+}) + v_{n-1} J_0(\alpha\theta_{-})] - \gamma_0 |v_n|^2 v_n.
\label{eq5}
\end{eqnarray}
Note that parameters $\gamma_1, \Omega \sim 1$, and the averaged equation is valid for times $t \leq 1/\epsilon$.
This  modified DNLS equation can be written as
$i \dot{v}_{n} = \delta H_{av}/\delta v_n^{\ast},$ with
averaged Hamiltonian
\begin{equation}
\label{ham}
H_{av}=-\sum_n\{\kappa J_0(\alpha\theta_{+}) \left[v_{n+1}v_n^{\ast}+v_{n+1}^{\ast}v_n \right]  +
\frac{\gamma_0}{2}|v_n|^4\}.
\label{eqq7}
\end{equation}
A comparison with Eq. (\ref{dnlsH}) gives the anticipated rescaling as  $\kappa \rightarrow \kappa J_0(\alpha\theta_{+})$; a similar rescaling was recently reported also for a  quantum Bose-Hubbard dimer with time dependent onsite interaction \cite{Gong}.

It is worth noting that while the appearance of the Bessel function is intimately connected with harmonic modulations, the existence of compacton solutions and the lattice tunneling suppression is generic for periodic SNLM. Thus, for example, for a two-step modulation of the form $\gamma(t) = (-1)^i\gamma_1$ with $i=0,1$ and $\frac i2 < \tau < \frac {(i+1)}2,\;$   we obtain for the first term in the averaged Hamiltonian (6)
$\kappa(v_{n+1}^{\ast}v_n e^{i\gamma_1\theta_+/4} + v_{n}^{\ast}v_{n+1} e^{-i\gamma_1\theta_+/4}){\rm sinc}(\gamma_1\theta_+/4)$,
where ${\rm sinc}(x)=\sin(x)/x$, thus, in this case the suppression of tunneling exists at zeros of the sinc function.
We also remark that for small $\alpha\theta_{+}$ the series expansion of  $J_0$
yields the averaged Hamiltonian of the DNLS equation
obtained in \cite{ATMK} in the limit of {\it weak} management.

{\it Exact compactons and numerics}.
To demonstrate the existence of {\it exact stable} compactons in the
averaged system, we seek for stationary solutions of the form
$v_n = A_n e^{-i \mu t}$ for which Eq. (\ref{eq5}) becomes
\begin{eqnarray}
\mu A_n + \gamma_0 A_n^3 +\kappa(A_{n+1}J_0(\alpha\theta_{+}) +
A_{n-1}J_0(\alpha\theta_{-})) + \nonumber\\
2\alpha\kappa A_n^2 [A_{n+1} J_1(\alpha\theta_{+}) +  A_{n-1} J_1
(\alpha\theta_{-})] = 0.
\label{eq9}
\end{eqnarray}
As is well known,  discrete  breathers can be numerically constructed with high precision using continuation procedures from the anti-continuous limit. The application of this method to Eq. (\ref{eq9}) gives, quite surprisingly, that
such modes {\it cannot}
be continued past a critical point (of $\kappa \approx 0.32$
for $-\mu=\gamma_0=1$). The fact that the solutions  cease to exist before reaching  the limit of resonance with the linear modes ($\kappa= -\mu/2$) naturally raises the question of what type of modes may be present in the system for larger values of the coupling. In the following we  show that in agreement with our theoretical prediction, the emerging  excitations  are genuine compactons e.g. they have vanishing tails (rather than fast double exponential decaying tails as in granular crystals~\cite{nesterenko}).
\begin{figure}
\centerline{
\includegraphics[width=4.5cm,height=6cm,clip]{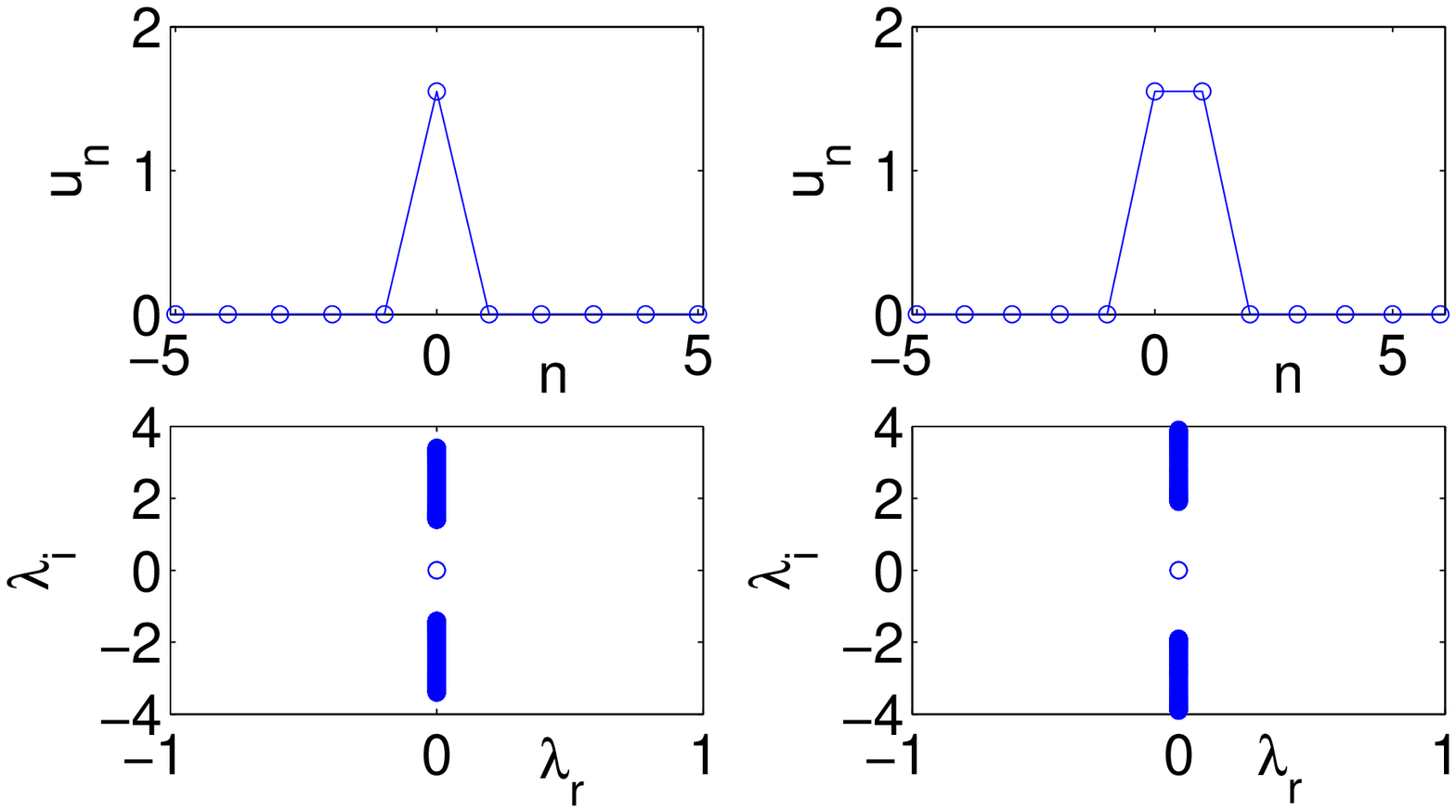}
\includegraphics[width=4.5cm,height=6cm,clip]{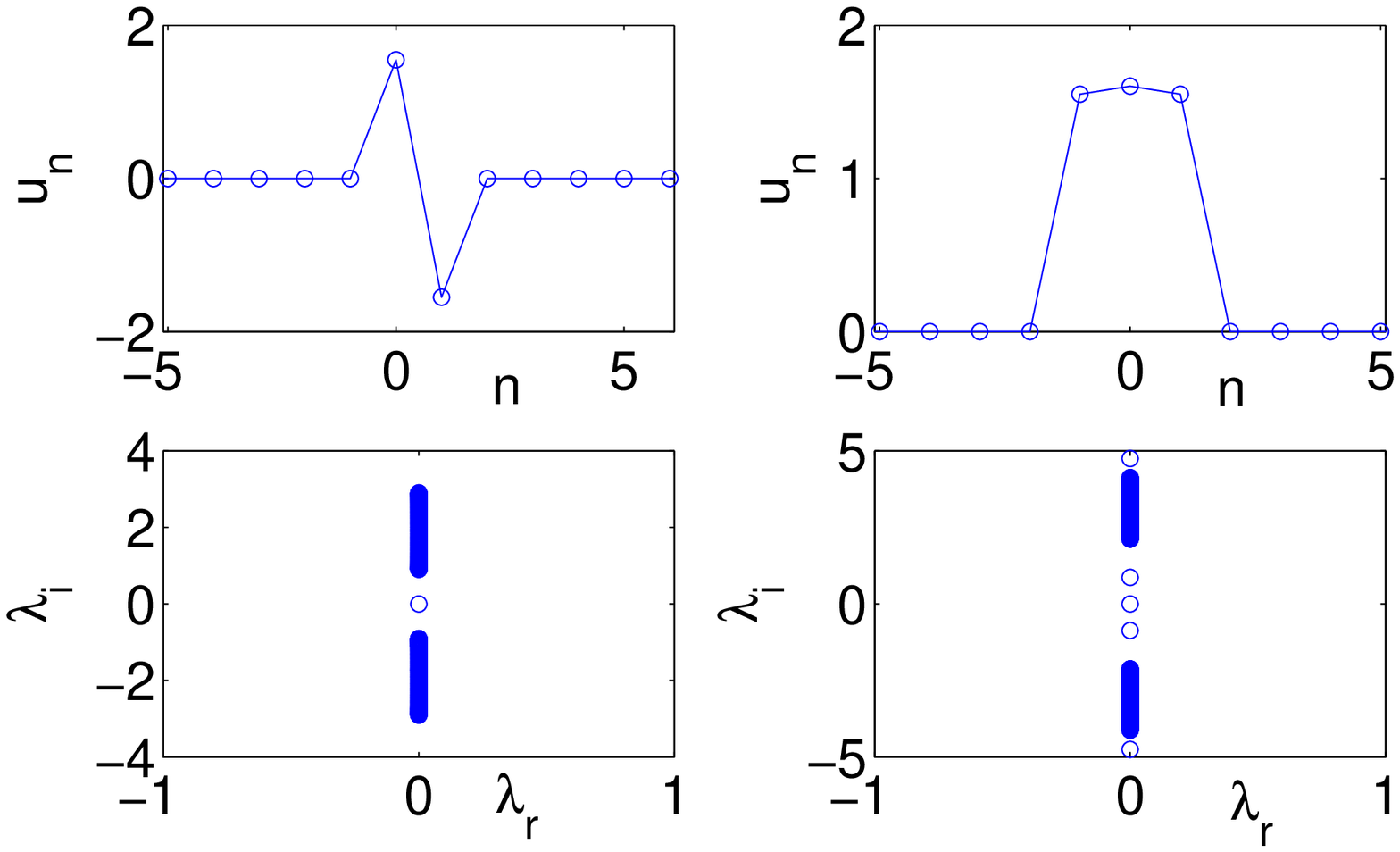}}
\caption{Typical examples for $\kappa=0.5$, $\alpha=1$ of
compact localized mode solutions of Eq. (\ref{eq9}) (top panels) and
of the plane $(\lambda_r,\lambda_i)$ of their
linearization eigenvalues $\lambda=\lambda_r + i \lambda_i$.
1st column: on-site, 2nd column: inter-site, in-phase, 3rd column:
inter-site, out-of-phase, 4th column: three-sites. Remarkably, all solutions
are {\it dynamically stable}.}
\label{fig2}
\end{figure}

To search for compactly supported solutions one needs to consider  \cite{konotop}
the last site of vanishing amplitude, denoted as $n_0$ below.
In the setting
of Eq. (\ref{eq9}), this directly establishes the condition
\begin{eqnarray}
J_0(\alpha A_{n_0+1}^2)=0 \Rightarrow A_{n_0+1}^2 = 2.4048/\alpha
\label{eq9a}
\end{eqnarray}
which yields the solution (based on the first zero of the Bessel
function) for the ``boundary'' of the compactly supported site.
Then, for $\mu=-\gamma_0
A_{n_0+1}^2$, both the condition for compact support
at $n_0 \pm 1$, and the equation for $n=n_0$ are satisfied. Hence
Eq. (\ref{eq9a}) yields a single-site discrete compacton.
\begin{figure}
\centerline{
\includegraphics[width=4.5cm,height=6cm,clip]{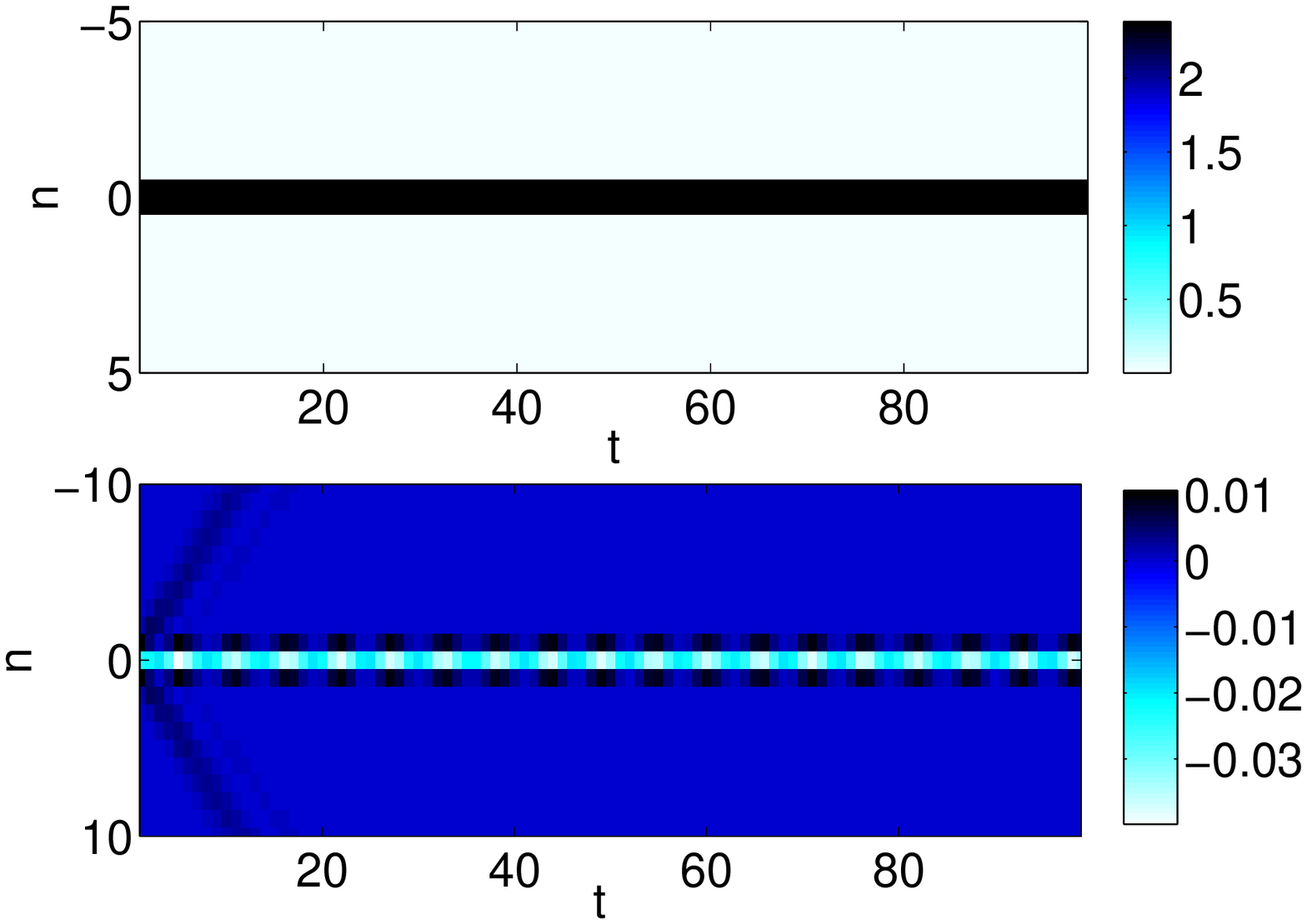}
\includegraphics[width=4.5cm,height=6cm,clip]{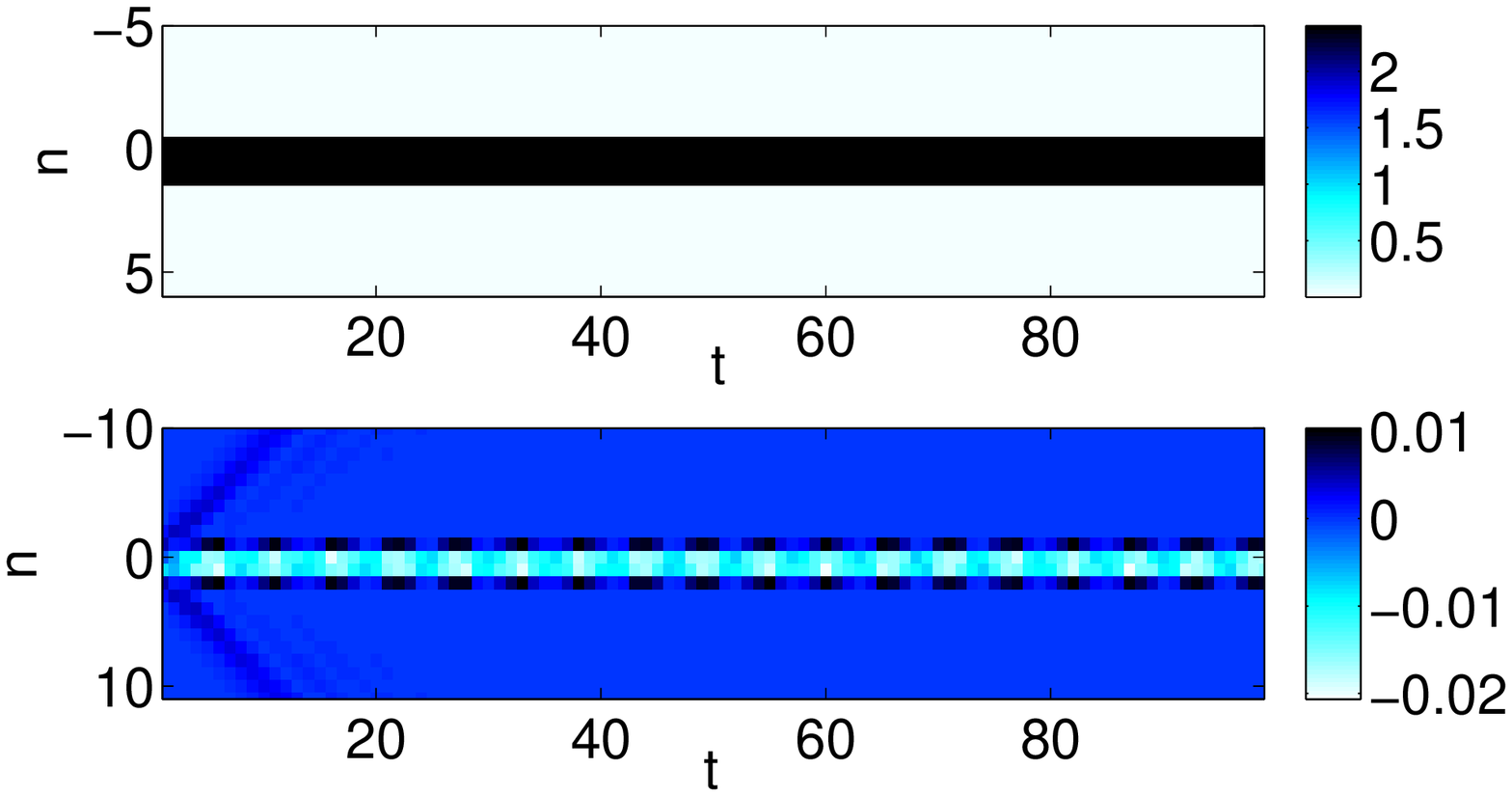}}
\caption{Space-time evolution of a one-site (left column) and a two-sites (right column) compacton solution as obtained from direct numerical integrations of Eq. (\ref{dnls}).
Top panels in each case show the square modulus of the solution itself (large
amplitude colorbar), while bottom panels (small amplitude colorbar)
show the deviation from the exact solution of Eq. (\ref{eq9}) taken as corresponding initial condition.}
\label{fig3}
\end{figure}
Numerical linear stability analysis
illustrates that this solution is {\it generically} stable
(see Fig. \ref{fig2}). The
bottom panel's eigenvalues are associated with perturbations growing
as $e^{\lambda t}$. The absence of a positive real part in $\lambda$
(i.e. of any $\lambda$'s in the right half plane) is tantamount to
linear stability.
Similar results are found for two-site compactons,
which are either in phase (2nd column of Fig. \ref{fig2})
or out-of-phase (3rd column of Fig. \ref{fig2}). The only thing that
changes here is that
in order to satisfy the equation at the non-vanishing sites, one must have
$\mu=-\kappa- \gamma_0 A_{n_0+1}^2,$  $\quad\mu=\kappa- \gamma_0 A_{n_0+1}^2$,
respectively for the in-phase and out-of-phase two-site compactons
(note from Fig. \ref{fig2} that  these solutions are both  stable).

With some additional effort, one can generalize these considerations to an arbitrary
number of sites. As a typical example, a three-sites compacton  with
amplitudes $(\dots,0,A_1,A_2,A_1,0,\dots)$ will satisfy in
addition to the ``no tunneling condition'' $J_0(\alpha A_1^2)=0$,
the constraints:
\begin{eqnarray}
&&
\mu A_{1+i} + 2 (i+1) \alpha \kappa
A_{1+i}^2 A_{2-i} J_1(\alpha (A_{2-i}^2-A_{1+i}^2))+
\nonumber \\ &&
\gamma_0 A_{1+i}^3 + \kappa A_{2-i} J_0(\alpha (A_{2-i}^2-A_{1+i}^2))
= 0,
\hspace{1mm}  i=0,1
\label{eq9d}
\end{eqnarray}
which can be easily solved to yield a solution as the one shown in
the 4th column of Fig. \ref{fig2}. We find that even such more
complex solutions
(which are highly unstable in DNLS \cite{pgk_book}) are
{\it dynamically robust} herein. This departure from the
standard DNLS model can be rationalized by the fact that in the latter
case the instability is mediated by the intersite
tunneling/coupling~\cite{pgk_book},
which for our
special compacton solutions vanishes, hence endowing the solutions
herein with dynamical stability.
\begin{figure}
\centerline{
\includegraphics[width=4.5cm,height=6cm,clip]{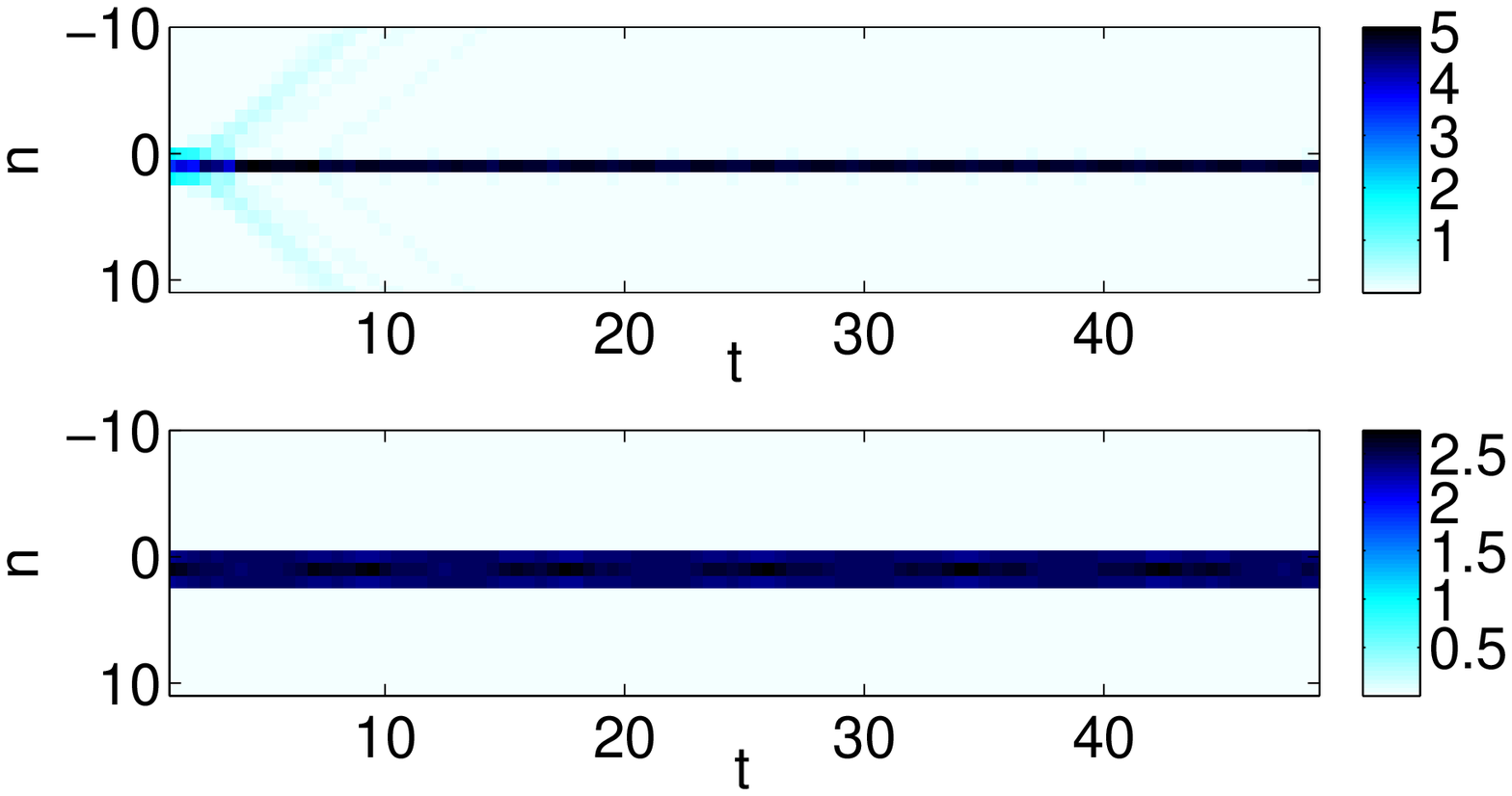}
\includegraphics[width=4.5cm,height=6cm,clip]{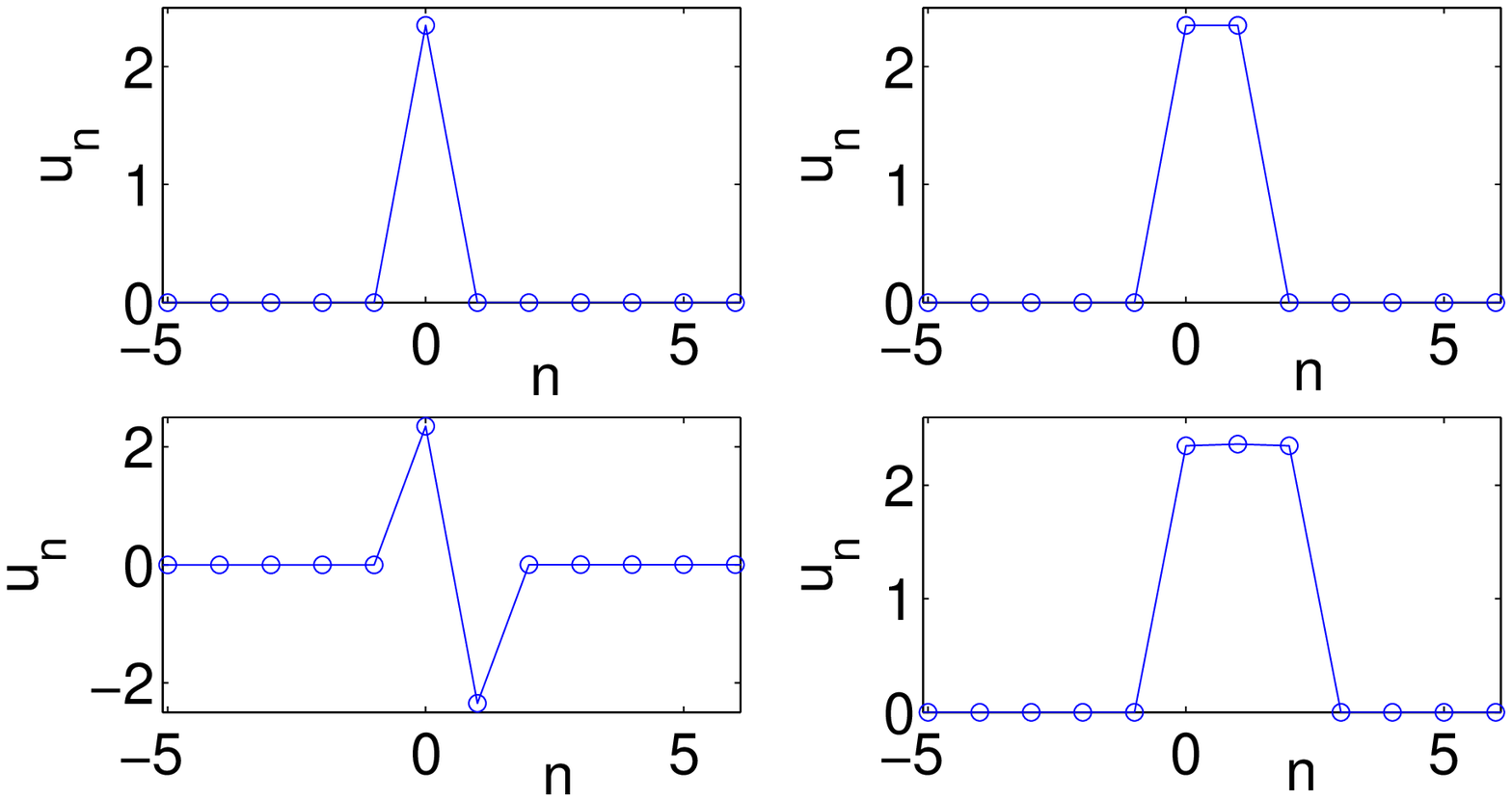}}
\caption{Left panel: dynamical evolution for $\kappa=1$
of a perturbed 3-sites compacton for $\epsilon=0.1$ (top, leading
to single-site evolution) and for $\epsilon=0.025$ (bottom).
Right panel: Examples of {\it stable} large amplitude compact modes with  one-, two- (in-phase and
out-of-phase) and 3-sites  emerging from the second zero
of the Bessel function.
}
\label{fig4}
\end{figure}

The dynamical stability of the solutions of Fig. \ref{fig2}
with respect to the original  DNLS model in Eq. (\ref{dnls}) has
been investigated  in
Fig. \ref{fig3} for
the one-site (left panels) and the
two-site, in-phase (right panels) modes (similar findings were
obtained for other modes). The top panels show
the space-time contour map of the solution modulus, while the bottom
panels
illustrates the deviation from the initial condition.
The structural stability of these compactons was ensured by adding
a uniformly distributed random perturbation
of small amplitude
to the original solution. Both for the averaged
equation (not shown here) and for the original system (see
Fig. \ref{fig3}), the
relevant perturbation stays bounded and never exceeds
$2 \%$ of the solution amplitude. The waveforms remain remarkably
localized in their compact shape (after a transient stage of shedding
off small amplitude ``radiation'')  and their tails never exceed
an O$(\epsilon)$ correction, as theoretically expected for
timescales of O$(1/\epsilon)$.
Notice that for Eq. (\ref{dnls}),
$\gamma(t)=1 +\frac{1}{\epsilon} \cos(t/\epsilon)$, with $\epsilon=0.1$
was used.
{\it However,} if one departs from the
regime of validity of the averaging and from the SNLM limit,
interesting deviations from the above behavior (and stability) arise.
An example
of this is shown in the left panels of Fig. \ref{fig4}.
In this case, the three-site solution was initialized in Eq. (\ref{dnls})
with $\epsilon=0.1$ in the top panel, while $\epsilon=0.025$
in the bottom one. In the latter, the above
argued robustness of the averaged modes was observed. Yet, in the
former one, the apparent lack thereof was clearly due to the
use of an $\epsilon$ outside of the regime of applicability of
the averaging approximation. Nevertheless, the resulting evolution
has two interesting by-products. Firstly, it confirms
the general preference of the system towards settling in
compact modes, since the evolution asymptotes to an
essentially single-site solution. Secondly, the larger
amplitude of this solution in comparison to those of Fig. \ref{fig2}
led us to explore the possibility of compactly supported modes
associated with {\it higher zeros} of the Bessel function in
the right panels of Fig. \ref{fig4}. Remarkably, such solutions
again, not only exist but are stable in all the
cases shown in the figure (numerical linear stability graphs are omitted).
This indicates the existence of an infinite sequence of such
modes, connected with the zeros of the Bessel structure  of the (averaged) tunneling.
\begin{figure}
\centerline{
\includegraphics[width=4.5cm,height=4.6cm,clip]{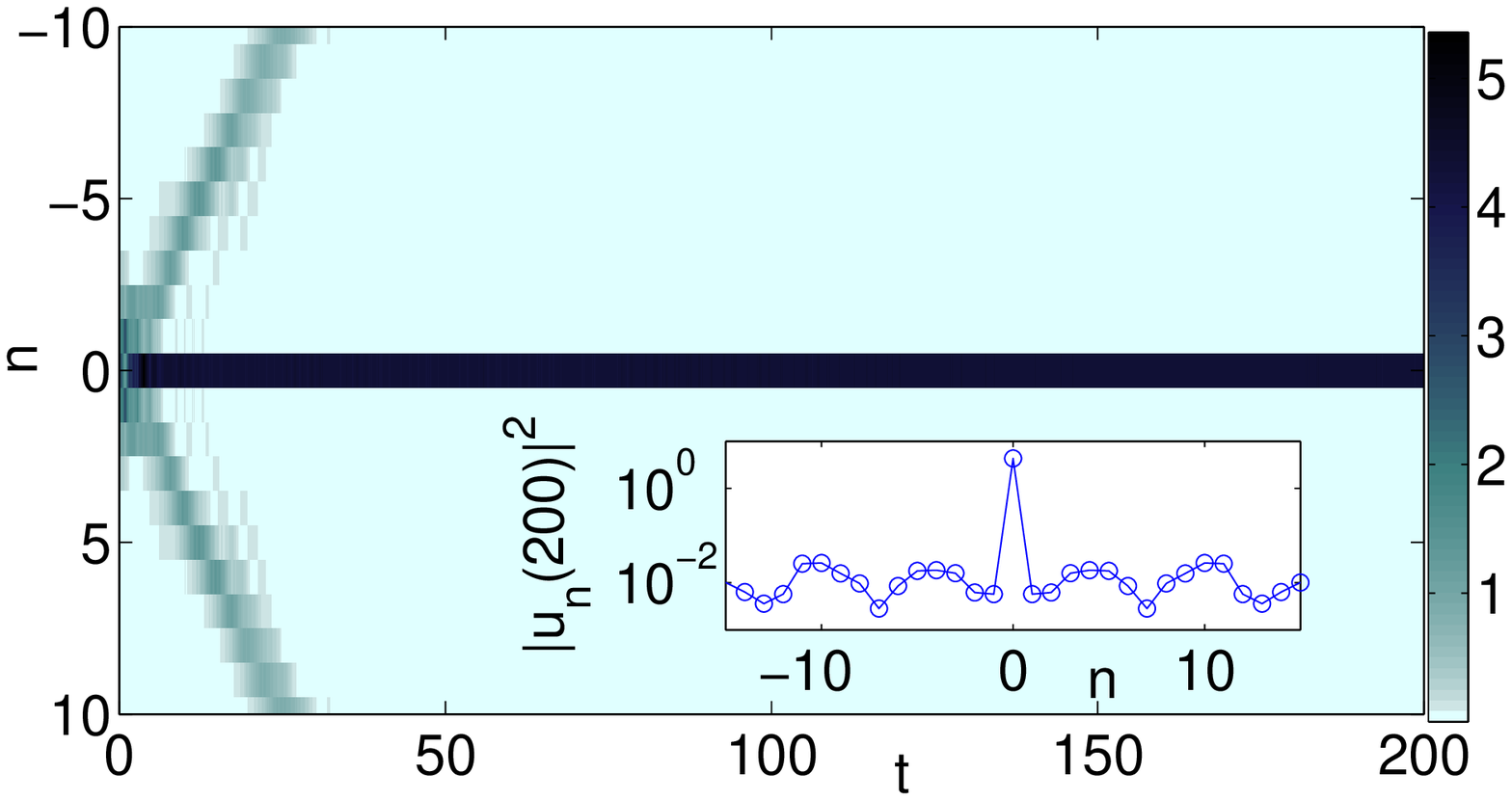}
\includegraphics[width=4.5cm,height=4.5cm,clip]{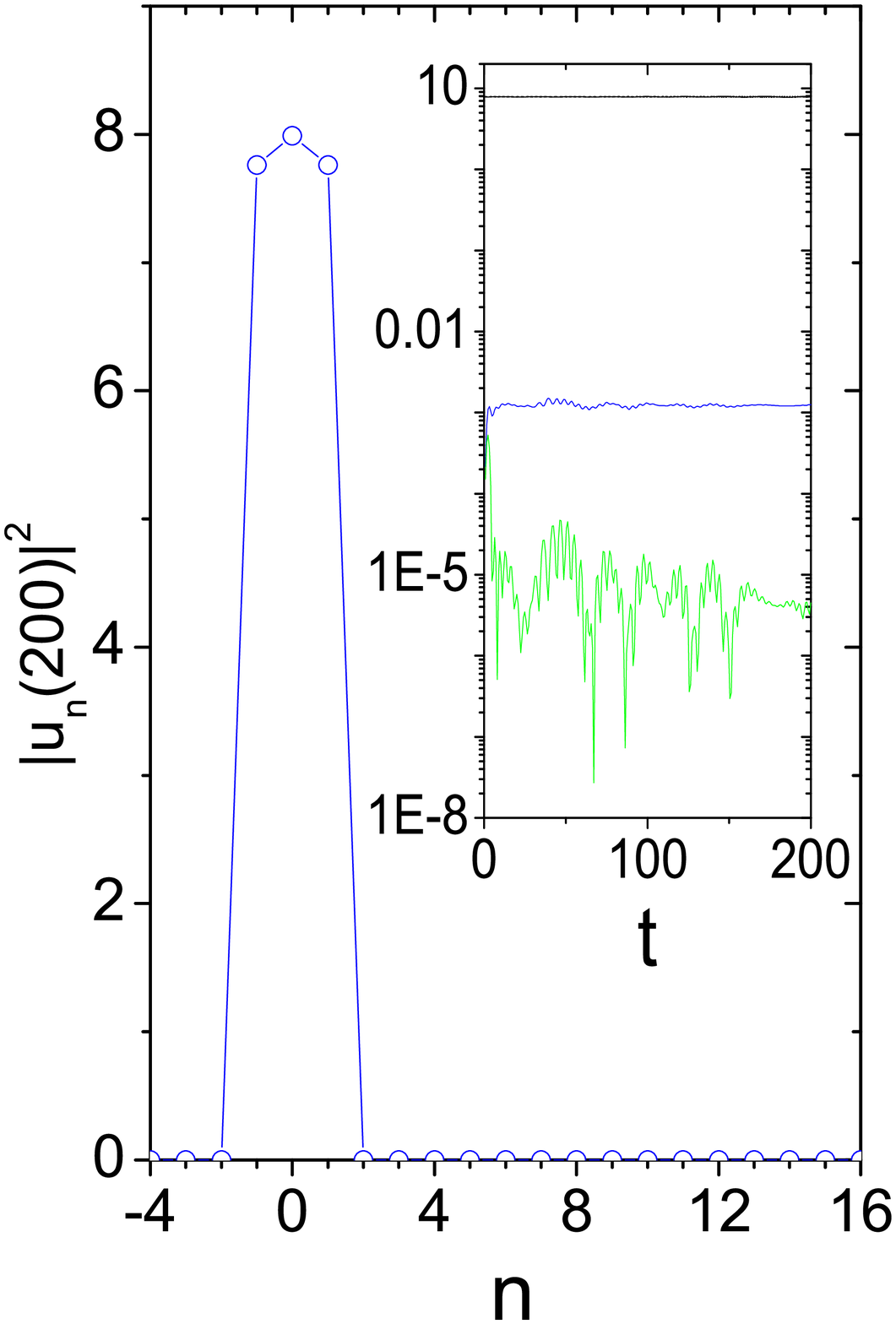}}
\caption{Left panel: Space-time evolution of a Gaussian wavepacket
$u_n(0)=1.5 e^{-0.1 n^2}$ under Eq. (\ref{dnls}) for $\kappa=0.5$. Clearly, an essentially
compact single-site excitation is produced (see the inset for $t=200$).
Right panel: Three-site compacton at time $t=200$ generated from uniform i.c. $u_i=2.8, i=-1,0,1$
for $\kappa=0.5$. The inset shows time evolution curves (from top to bottom) of amplitudes
at $0,1,2,3$, respectively (the first two curves overlap in this scale).   }
\label{fig5}
\end{figure}

To address the robust emergence of such compact excitations,
we used both few-sites uniform and even Gaussian type excitations.
While in the former (experimentally realizable, see e.g.
\cite{silber}) case, discrete compactons can be expected,
remarkably, in either case such excitations can result.
A typical example is shown for a Gaussian initial profile
in the left panel of  Fig. \ref{fig5}, which yields a single-site
compact mode differing
in amplitude by more than two orders of magnitude between the central site and
its nearest neighbor and showing {\it no signs whatsoever of an exponential
tail}, even in a semilog plot. In the right panel of Fig. \ref{fig5} a
multi-site compacton generated from uniform compactly supported data is
depicted. We see that the amplitude reduces by five orders of magnitude 3-sites away from the central peak.

Let us estimate the parameters for the experimental observation of such
modes,
e.g. for the case of $^7$Li condensate in a deep optical lattice.  The
Feshbach resonance  in Li occurs at the value of external magnetic field
$B=720 G$. By varying the magnetic field around this value we can easily
obtain variations of the scattering length $a_{s1}$ around the
order of the  background scattering length $a_{s0}$
yielding $\gamma_1/\gamma_0 \sim 10$.  In the deep optical lattice with
$V_0 > 10 E_R$, where
 $V_0$ is the depth of the lattice and $E_R = \hbar^2 k^2/2m$ is the recoil
energy, the Gross-Pitaevskii equation can be mapped into the DNLS
equation (\ref{dnls}) \cite{ABKS}.
Thus by changing periodically in time the magnetic field between these
values  with the frequency $\Omega \sim 10 \omega_{R}$,
where $\omega_R=E_R/\hbar$, we can generate matter wave compactons.

\noindent {\it Conclusions}. We predicted the existence of discrete compactons in
the DNLSE with strong nonlinearity management.
We found stable  single and few-sites compactons of odd and even parity.
They are
 robust and can be generated from different classes of
initial  conditions. Such structures may be observable
in experiments on BECs in deep optical lattices with periodically varying
scattering length and arrays of nonlinear optical waveguides with variable
Kerr coefficient along the propagation distance.

FKA acknowledges the European Community for two years grant PIIF-GA-2009-236099. PGK acknowledges support from NSF-DMS-0349023,
NSF-DMS-0806762 and the A. von Humboldt Foundation.
MS thanks the MIUR for support through  a PRIN-2008 initiative.

\end{document}